\documentclass[published]{JHEP3}
\JHEP{10(2003)033}
\received{August 13, 2003}
\accepted{October 14, 2003}
\keywords{Topological Field Theories, Duality in Gauge Field Theories, Gauge Symmetry, Quantum Groups}

\usepackage{amsmath}

\newcommand\gl{\mathfrak}
\newcommand\dl{\mathbb}

\def\tilde{\widetilde}
\def\hat{\widehat}
\newcommand{\NR}{{{\dl R}}}
\newcommand{\NC}{{{\dl C}}}
\newcommand{\tr}{\mathop{\rm tr}}
\newcommand{\ad}{\hbox{ad}}
\newcommand{\CC}{{\cal C}}
\newcommand{\CL}{{\cal L}}

\newcommand{\Ad}{{\rm Ad}} 
\newcommand{\gp}{{g_{_+}}} 
\newcommand{\gm}{{g_{_-}}} 
\newcommand{\gpinv}{{g_{_{+}}^{-1}}}
\newcommand{\gminv}{{g_{_{-}}^{-1}}}

\title{Topological Poisson sigma models on Poisson-Lie groups}

\author{Iv\'an Calvo, Fernando Falceto and David
Garc\'{\i}a-\'Alvarez\\ 
Departamento F\'{\i}sica Te\'orica, Univ.\ Zaragoza\\
E-50009 Zaragoza, Spain\\
E-mail: \email{icalvo@unizar.es}, 
\email{falceto@unizar.es}, \email{dga@saturno.unizar.es}}

\abstract{We solve the topological Poisson Sigma model for a
Poisson-Lie group $G$ and its dual $G^*$.  We show that the gauge
symmetry for each model is given by its dual group that acts by
dressing transformations on the target. The resolution of both models
in the open geometry reveals that there exists a map from the reduced
phase space of each model ($P$ and $P^*$) to the main symplectic leaf of the
Heisenberg double ($D_{0}$) such that the symplectic forms on $P$,
$P^{*}$ are obtained as the pull-back by those maps of the symplectic
structure on $D_{0}$.  This uncovers a duality between $P$ and $P^{*}$
under the exchange of bulk degrees of freedom of one model with
boundary degrees of freedom of the other one. We finally solve the
Poisson Sigma model for the Poisson structure on $G$ given by a pair
of $r$-matrices that generalizes the Poisson-Lie case. The hamiltonian
analysis of the theory requires the introduction of a deformation of
the Heisenberg double.}

\begin{document} 

\section{Introduction}\label{section1}

Since their appearance in~\cite{Strobl,Ikeda} the Poisson Sigma
models have played a central role in the study of two dimensional
gauge theories.  They are related to pure gravity, to
Wess-Zumino-Witten models and Yang-Mills in two dimensions.
Generalizations of the model that include supergravity have been
recently proposed in~\cite{EKS,BK}.

The Poisson Sigma models are topological and besides 
the two dimensional (space-time) manifold $\Sigma$ 
all we require is a Poisson structure on the target $(M,\{\ ,\})$
(supplemented by boundary conditions if 
$\partial\Sigma\not=\emptyset$). 
The theory seems then suitable for the
study of the underlying Poisson structure and in fact 
in~\cite{CaFe} it was shown that
the semiclassical expansion of these models
(with $\Sigma$ the unit disk and insertions at the boundary)
reproduces the $*$-product introduced by Kontsevich
in~\cite{Kon} that quantizes by deformation the Poisson
bracket.

In this paper we shall be interested in Poisson Sigma models in which
$M$ is a Lie group.  In this case it is natural to ask for a
consistency relation between its Poisson and group structures and this
leads to the concept of Poisson-Lie group (\cite{STS,Lu}).
Poisson-Lie groups are the classical counterparts of quantum groups
and they lead to the so called \emph{hidden symmetries} of integrable
systems that induce dressing transformations in the phase space. We
shall give later a more detailed description of these objects, but for
the moment we would like to stress that Poisson-Lie groups come in
dual pairs under the exchange of the roles of Poisson and Lie
brackets.  Examples of Poisson-Lie Sigma models have been studied
in~\cite{Falceto,ASS} in connection to $G/G$ Wess-Zumino-Witten
theories. Here we pursue a systematic study of the matter and besides
we shall also consider some generalizations of the Poisson structure
for which the product with the corresponding Poisson-Lie group acts as
a Poisson action.

\pagebreak[3]

One of the simplest examples of Poisson-Lie Sigma models is
the linear one. Here $M$ is a vector space (abelian group) and its
Poisson structure is linear. This model is related to $BF$ and 
Yang-Mills  and can be considered as dual of the trivial Poisson-Lie
Sigma model in an (in general) non-abelian group with vanishing
Poisson bracket.  Our models can be regarded as the simplest 
generalizations of the linear ones with which they share some 
properties that will be stressed in the sequel.

Another aspect that will deserve our attention is that of duality,
i.e.\ we shall try to relate the two dual models, which is not obvious
at the lagrangian level. However, in the hamiltonian approach, in the
open geometry, the duality becomes evident: it consists of the
exchange of bulk and boundary degrees of freedom. We would like to
relate this fact with the non-abelian T-duality for WZW Sigma models
(\cite{Klimcik}).

We begin with a brief review of the essential results about Poisson
Sigma models and Poisson-Lie groups in sections~\ref{PSmodels}
and~\ref{PoissonLiegroups}. Sections~\ref{G} and~\ref{Gdual} are
devoted to the study of the Poisson-Lie Sigma model of a simple
Poisson-Lie group and its dual one, respectively.  By using the
techniques of the two previous sections we are able to solve more
general Poisson Sigma models on Lie groups and we introduce a
generalization of the Heisenberg double in
section~\ref{generalization}. The conclusions and open questions are
presented in section~\ref{conclusions}.

\section{Poisson Sigma models} \label{PSmodels}\label{section2}

The Poisson Sigma model is a two-dimensional topological Sigma model
with the target manifold $M$ equipped with a Poisson structure
$\Gamma\in T^{\wedge 2}(M)$, i.e.\ a bivector field on $M$. The
Poisson bracket of two functions on $M$ is given by the contraction of
$\Gamma$: $\{f,g\}(m)=\iota(\Gamma_m) df\wedge dg$.

The fields of the model are $X:\Sigma \rightarrow M$ and a 1-form
$\psi$ on $\Sigma$ with values in the pullback by $X$ of the cotangent
bundle of $M$.  The action functional has the form
\begin{equation}  \label{PS}
S_{P\sigma}(X,\psi)=\int_\Sigma \langle dX,\wedge\psi\rangle - 
 \frac 12 \langle\Gamma\circ X,\psi\wedge\psi\rangle
\end{equation} 
where $\langle,\rangle$ denotes the pairing between the tangent and
the cotangent vectors to $M$.

If $X^{i}$ are local coordinates on $M$, $\sigma ^{\mu}$ local
coordinates on $\Sigma$, $\Gamma^{ij}$ the components of the Poisson
structure in these coordinates and $\psi_{i}=\psi_{i\mu}d{\sigma}^{\mu}$,
$i=1,\ldots ,n$, $\mu=1,2$ the action reads
\begin{equation} 
\label{PScoor}
S_{P\sigma}(X,\psi)=\int_\Sigma dX^{i}\wedge \psi_{i}-
 \frac 12 \Gamma^{ij}(X)\psi_{i}\wedge \psi_{j}
\end{equation} 

It is straightforward to work out the equations of motion:
\begin{subequations}
\begin{eqnarray}
dX^{i}+\Gamma^{ij}(X)\psi_{j}&=&0 
\label{eomA}
\\
d\psi_{i}+ \frac 12 \partial_{i}\Gamma^{jk}(X)\psi_{j}\wedge \psi_{k}&=&0 
\label{eomB}
\end{eqnarray}
\end{subequations}

One can show (\cite{BojoStrobl}) that for solutions
of~(\ref{eomA}) the image of $X$ lies within 
one of the symplectic leaves  of the foliation of $M$.

Under the infinitesimal transformation
\begin{eqnarray} 
\label{symmetry}
\delta_{\epsilon}X^{i}&=&\epsilon_{j}\Gamma^{ji}(X)
\nonumber\\ 
\delta_{\epsilon}\psi_{i}&=&d\epsilon_{i}+\partial_{i}\Gamma^{jk}(X)\psi_{j}
\epsilon_{k}
\end{eqnarray}
where $\epsilon=\epsilon_{i}dX^i$ is a section of $X^*(T^*(M))$, the
action~(\ref{PScoor}) transforms by a boundary term
\begin{equation}  
\label{symmS}
\delta_{\epsilon}S_{P\sigma}=-\int_\Sigma d(dX^i \epsilon_i)\,.
\end{equation} 

One of the special properties of these models is that the commutator
of two consecutive gauge tansformation of~(\ref{symmetry}) is not of
the same form i.e.\ 
\begin{subequations}
\begin{eqnarray}
 [\delta_\epsilon,\delta_{\epsilon'}]X^i&=&
\delta_{[\epsilon,\epsilon']^*} X^i
\label{commgA}
\\{}
[\delta_\epsilon,\delta_{\epsilon'}]\psi_i&=&
\delta_{[\epsilon,\epsilon']^*} \psi_i
+\epsilon_k\epsilon_{l}'\partial_i\partial_j
\Gamma^{kl}(dX^{j}+\Gamma^{js}(X)\psi_{s})
\label{commgB}
\end{eqnarray}
\end{subequations}
where for a vector field $\chi\in{\gl X}(M)$,
$\iota(\chi)[\epsilon,\epsilon']^*= \iota(\CL_\chi
\Gamma)\epsilon\wedge\epsilon'$. Later, we shall make extensive use of
this commutator in the context of Poisson-Lie groups.  Note that the
term in parenthesis in~(\ref{commgB}) is the equation of
motion~(\ref{eomA}), and then it vanishes on-shell. However, as
remarked in~\cite{BojoStrobl}, due to the $X$ dependence of
$[\epsilon,\epsilon']^*$ even on-shell one can not talk properly about
a Lie algebra structure in the space of parameters, unless we enlarge
its definition to account for field dependent gauge
transformations. The situation will become simpler in the linear case
that we describe below.

One of the simplest examples of Poisson Sigma models is the linear
one (2D BF theory).  We take a finite dimensional vector space $V$ as the target
space and define a linear Poisson bracket on it, i.e.\ for any
$\phi,\psi\in V^*$, their Poisson bracket $\{\phi,\psi\}\in V^*$.  The
linear Poisson bracket on $V$ defines then a Lie bracket $[\ ,\ ]$ on
$V^*\equiv{\gl g}$. Assuming that $\gl g$ is semisimple we can use the
Cartan-Killing form $\tr$ to identify $\gl g$ and ${\gl g}^*$.  The
fields are then $X\in \Lambda^0(\Sigma)\otimes{\gl g}$ and $A\in
\Lambda^1(\Sigma)\otimes{\gl g}$ and the action reads:
\begin{equation}
\label{linac}
S_{l}(X,A)=\int_\Sigma \tr \left(dX\wedge A+ \frac 12 A\wedge
[X,A]\right)
\end{equation} 

The equations of motion are:
\begin{eqnarray}
\label{lineq}
dA+[A,A]&=&0
\nonumber\\
dX+[A,X]&=&0
\end{eqnarray}

For $\epsilon\in \Lambda^0(M)\otimes{\gl g}$ the gauge transformation
\begin{eqnarray}
\delta_\epsilon X&=& [X,\epsilon]
\nonumber\\
\delta_\epsilon A&=& d\epsilon+ [A,\epsilon]
\end{eqnarray}
induces the change of the action~(\ref{linac}) by a boundary term
\begin{equation} 
\delta_\epsilon S_{l}=-\int_\Sigma d\tr (dX\epsilon)\,.
\end{equation} 

Note that in this case the gauge transformations close even off-shell
$$
[\delta_\epsilon,\delta_{\epsilon'}]=\delta_{[\epsilon,\epsilon']}
$$
and induce the Lie algebra structure of $\Lambda^0(\Sigma)\otimes{\gl
g}$ in the space of parameters.

\section{Poisson-Lie groups} \label{PoissonLiegroups}\label{section3}

In this section we shall review some basic features of the theory of
Poisson-Lie groups and fix our
notation. See~\cite{STS,Falceto,Lu,Alekseev,Falceto2} for details.  A
Poisson-Lie group is a Lie group equipped with a Poisson structure
which makes the product $m:G\times G \rightarrow G$ a Poisson map if
$G \times G$ is considered with the product Poisson
structure. Linearization of the Poisson structure at the unit $e$ of
$G$ provides a Lie algebra structure on ${\gl g}^*=T^*_{e}(G)$ by the
formula
\begin{equation} 
\label{Liedual}
[d\xi_{1}(e),d\xi_{2}(e)]^{*}=d\{\xi_{1},\xi_{2}\}(e)
\end{equation}

The Poisson-Lie structure of $G$ yields the compatibility condition
\begin{equation} 
\label{comp} 
\left< [X,Y],[v,w]^*\right>+\left<\ad_v^*X,\ad_Y^*w\right>
-\left<\ad_w^*X,\ad_Y^*v\right>
-\left<\ad_v^*Y,\ad_X^*w\right>+\left<\ad_w^*Y,\ad_X^*v\right>=0\,.
\end{equation} 
which allows to define a Lie bracket in ${\gl g}\oplus {\gl g}^*$ by
the formula
\begin{equation} 
\label{Liesuma}
[X+\xi,Y+\zeta]=[X,Y]+[\xi,\zeta]^{*}-ad^{*}_{X}\zeta + ad^{*}_{Y}\xi
+ ad^{*}_{\zeta}X - ad^{*}_{\xi}Y
\end{equation}

If $G$ is connected and simply connected,~(\ref{comp}) is enough to
integrate $[\ ,\ ]^*$ to a Poisson structure on $G$ that makes it
Poisson-Lie and the Poisson structure is unique. The symmetry between
${\gl g}$ and ${\gl g}^*$ in~(\ref{comp}), implies that one has also a
Poisson-Lie group $G^*$ with Lie algebra $({\gl g}^*,[\ ,\ ]^*)$ and a
Poisson structure whose linearization at $e$ gives the bracket $[\ ,\
]$. $G^*$ is the dual Poisson-Lie group of $G$.

Let us take $G$ a complex, simple, connected, simply connected Lie
group and give the above construction explicitly. The (essentially
unique) nondegenerate, invariant, bilinear form tr on ${\gl g}$
establishes an isomorphism between ${\gl g}$ and ${\gl g}^*$. The
Poisson structure $\Gamma$ contracted with the right-invariant forms
$\Lambda (X)=\tr(dgg^{-1}X),\ X \in {\gl g}$, will be denoted
$$
\gamma_{g}(X,Y)=\iota(\Gamma_{g})\Lambda(X)\wedge\Lambda(Y)
$$

For a general Poisson-Lie structure on $G$ (\cite{Lu}),
\begin{equation} 
\label{PoissonLie}
\gamma_{g}^{r}(X,Y)= \frac 12 \tr(XrY-XAd_{g}rAd_{g}^{-1}Y)
\end{equation}
where $r:{\gl g}\rightarrow {\gl g}$ is an antisymmetric linear
operator such that
\begin{equation} \label{YangBaxter}
r[rX,Y]+r[X,rY]-[rX,rY]=\alpha[X,Y]\,,\qquad \alpha \in \NC
\end{equation} 

Using the operator $r$ it is possible to define a new Lie bracket in
${\gl g}$,
\begin{equation}  \label{rproduct}
[X,Y]_{r}= \frac 12 ([X,rY]+[rX,Y])
\end{equation} 
$({\gl g},[\ ,\ ]_{r})$ is isomorphic to $({\gl g}^{*},[\ ,\ ]^{*})$
via tr.

If $\alpha\neq 0$ we can rescale the bilinear form in ${\gl g}$ and
$r$ so that we leave the Poisson bracket unchanged but $\alpha = 1$.
In the following we shall consider this case ($\alpha=1$) that
corresponds to the factorizable Lie bialgebras of ref.~\cite{STS2}.
Take $r_{\pm}= \frac 12 (r\pm I)$, ${\gl d}={\gl g}\oplus{\gl g}$ and
$[\ ,\ ]_{{\gl d}}=([\ ,\ ],[\ ,\ ])$.  Then,
\begin{equation}  
\label{homomorphism}
[r_{\pm}X,r_{\pm}Y]=r_{\pm}[X,Y]_{r}
\end{equation} 
and the embedding $\sigma: {\gl g}\rightarrow {\gl d}, X \mapsto 
(r_{+}X,r_{-}X)$ defines an homomorphism from $({\gl g},[\ ,\ ]_{r})$ 
to $({\gl d},[\ ,\ ]_{\gl d})$. We give an invariant, nondegenerate 
bilinear form on ${\gl d}$ by
\begin{equation}  
\label{traceond}
Tr((X_{1},X_{2})(Y_{1},Y_{2}))=\tr(X_{1}Y_{1})-\tr(X_{2}Y_{2})
\end{equation} 

This allows to identify $({\gl g}^{*},[\ ,\ ]^{*})$ with the
subalgebra ${\gl g}_{r}:=\sigma ({\gl g})\subset{\gl d}$ and $G^{*}$
as the subgroup $G_{r}$ in $D:=G \times G$ corresponding to the Lie
subalgebra ${\gl g}_{r}$.  We denote by $(g_{+},g_{-})$ the elements
of $G_{r}$. In a natural way, ${\gl g}\cong {\gl g}_{d}:=\{(X,X)\in
{\gl d}\mid X\in{\gl g}\}$ and $G\cong G_{d}:= \{(g,g)\in D\mid g\in
G\}$.  Notice that ${\gl g}_{d}\bigcap{\gl g}_{r}=0 \Rightarrow
G_{0}:= G_{d}\bigcap G_{r}$ is a discrete subgroup of $G$.

The induced Poisson-Lie structure on $G_{r}$ contracted with the
right-invariant forms on $G_{r}$,
$\Lambda^{r}(X)=tr[(dg_{+}g_{+}^{-1}- dg_{-}g_{-}^{-1})X]$ for $X \in
{\gl g}$, takes the form
\begin{equation} 
\label{dualPoissonLie}
\gamma_{(g_{+},g_{-})}^{r}(X,Y)=\tr\left[X(Ad_{g_{+}}-Ad_{g_{-}})
(r_{-}Ad_{g_{+}}^{-1}-r_{+}Ad_{g_{-}}^{-1})Y\right].
\end{equation} 

The $r$-matrices are used not only to give explicit realizations of
the dual group and to construct Poisson-Lie structures but to
construct more general Poisson structures on Lie groups. Following
Semenov-Tian-Shansky (\cite{STS}), if $r$ and $r'$ are antisymmetric
and satisfy~(\ref{YangBaxter}) with the same value for $\alpha$, we
can define the Poisson structure (contracted with the right-invariant
forms),
\begin{equation}  \label{generalPoisson}
\gamma_{g}^{r,r'}(X,Y)= \frac 12 \tr(XrY+XAd_{g}r'Ad_{g}^{-1}Y)
\end{equation} 

$G$ equipped with~(\ref{generalPoisson}) is denoted by $G_{r,r'}$.
One can verify that $G_{r,-r'}\times G_{r',r''}\rightarrow G_{r,r''}$
is a Poisson map. When $r'=r$ and $r''=-r$ then we get~(\ref{PoissonLie})
and the Poisson structure is, indeed, Poisson-Lie.

Given a $r$-matrix in $G$ it is possible to define a $r$-matrix in
$D=G \times G$ (which we shall denote by $R$) in a natural way: $R=
P_{d}-P_{r}$ where $P_{d}$ and $P_{r}$ are the projectors on ${\gl g}$
and ${\gl g}_{r}$ respectively, parallel to the complementary
subalgebra. Hence, $R_{+}=P_{d},\ R_{-}=-P_{r}$.

We shall need the description of the Heisenberg double (which is not a
Poisson-Lie group; in fact, with the notation introduced before, it
corresponds to $D_{R,R}$).  Its main symplectic leaf is
$D_{0}=G_{d}G_{r}\bigcap G_{r}G_{d}$ (that contains a neighborhood of
the unit of $D$). The symplectic structure obtained by inverting the
Poisson structure on it reads
\begin{equation} \label{Hsymp}
\Omega((h_{+}g,h_{-}g))=
\tr\left[d\tilde{g}\tilde{g}^{-1}\wedge 
(dh_{+}h_{+}^{-1}-
dh_{-}h_{-}^{-1})+g^{-1}dg\wedge (\tilde{h}_{+}^{-1}d\tilde{h}_{+}-
\tilde{h}_{-}^{-1}d\tilde{h}_{-})\right]
\end{equation} 
where $(h_{+}g,h_{-}g)=({\tilde g}{\tilde h}_{+}, {\tilde g}{\tilde
h}_{-})\in D_{0}$. Note that although these decompositions are not
unique, the ambiguity (the product by an element of $G_0$) does not
affect the values of $\Omega$ which is well defined in $D_0$.

\section{Poisson-Lie Sigma model on $G$} \label{G}\label{section4}

Recall the action~(\ref{PS}) and take the Poisson-Lie group of the
previous section as the target.  The action for $g:\Sigma\rightarrow
G$ and $A\in\Lambda^1(\Sigma)\otimes{\gl g}$ reads:
\begin{equation} 
\label{PLS}
S_{\rm PL}(g,A)=\int_{\Sigma}{\tr(dgg^{-1}\wedge A)-\frac 14
\tr\left(A\wedge (r-Ad_{g}rAd_{g}^{-1})A\right)}
\end{equation}
which is what we call the Poisson-Lie Sigma model for $G$.

\pagebreak[3]

We proceed now to a detailed study of this model, from which we shall
learn techniques applicable to the resolution of more general Poisson
Sigma models on Lie groups.

The equations of motion are:
\begin{eqnarray} 
dgg^{-1}+ \frac 12 (r-Ad_{g}rAd_{g}^{-1})A&=&0
\label{eomPLSA}
\nonumber\\
d{\tilde A}+[{\tilde A},{\tilde A}]_{r}&=&0\,,\qquad 
{\tilde A}:=Ad_{g}^{-1}A
\label{eomPLSB}
\end{eqnarray}

From the previous two equations one can derive also a zero 
curvature equation for $A$ itself
\begin{equation}\label{Aflat}
dA+[{A},{A}]_{r}=0\,,
\end{equation}
which can be deduced directly from the stationary condition for the
action under variations of the fields that keep $\tilde A$ invariant.

The  gauge symmetry of the action, for 
$\beta:\Sigma\rightarrow {\gl g}$, is given in its infinitesimal form 
by\begin{eqnarray} 
\delta_{\beta}g g^{-1}&=& \frac 12 (Ad_{g}rAd_{g}^{-1}-r)\beta 
\label{infsymmetryA}
\nonumber\\
\delta_{\beta}A&=&d\beta +
    [A,\beta]_{r}- \frac 12 \left[dgg^{-1}+ \frac 12 
(r-Ad_{g}rAd_{g}^{-1})A,\beta\right].
\label{infsymmetryB}
\end{eqnarray}

Under this transformation the action changes by a boundary term,
namely $\delta_{\beta}S_{\rm PL}=\int_\Sigma d\tr(g dg^{-1} \beta).$
Note that~(\ref{infsymmetryA}) corresponds to the right dressing
vector fields of ref.~\cite{Lu} translated to the origin by right
multiplication in $G$. Its integration (local as in general the vector
field is not complete) gives rise to the dressing transformation of
$g$. Also in~(\ref{infsymmetryB}) the third term vanishes on-shell; it
corresponds to a \emph{trivial} transformation (see
refs.~\cite{BojoStrobl,TeHe}), proportional to the equations of
motion. If we forget about that term (or alternatively, working
on-shell)
$$
[\delta_{\beta_{1}},\delta_{\beta_{2}}]=
\delta_{[\beta_{1},\beta_{2}]_{r}}\,.
$$
Note that unlike the general case this relation defines a Lie algebra
structure (that of the gauge group corresponding to $G_r$) in the
space of gauge transformations. This is also what happens in the
linear case discussed in section~\ref{section2}. However, in this case
the relation holds uniquely on-shell while in the linear case the Lie
algebra structure persists off-shell. Another difference is that the
vector fields (on-shell) induced by the gauge transformation are not
complete in the general case. This is a second obstruction to properly
talk about a gauge group, so we shall restrict ourselves to
infinitesimal gauge transformations.

Now we proceed to integrate the equations of motion. We shall consider
$\alpha=1$ in eq.~(\ref{YangBaxter}) although as remarked before the
case $\alpha\not=0$ can be treated similarly.  As the equation of
motion for $A$ is independent of $g$, like in the linear model, we
first solve the eqs.~(\ref{eomPLSB}) and~(\ref{Aflat}). This is in
contrast with the general case of ref.~\cite{BojoStrobl}. Locally
\begin{eqnarray}
A&=&h_{+}dh_{+}^{-1}-h_{-}dh_{-}^{-1}  
\label{solutionAA} 
\nonumber\\
{\tilde A}&=& {\tilde h}_{+}d{\tilde h}_{+}^{-1}-{\tilde
  h}_{-}d{\tilde h}_{-}^{-1} 
\label{solutionAB}
\nonumber\\ 
&&(h_{+},h_{-})\ {\rm and}\ ({\tilde h}_{+},{\tilde h}_{-}):U
\rightarrow G_{r}\,,\qquad 
\Sigma \supset U \ \mbox{open contractible.}
\end{eqnarray}

In order to fix the ambiguity in the new fields we choose a point
$\sigma_0\in U$ and take $h_\pm(\sigma_0)=\tilde h_\pm(\sigma_0)=e$.

We need to work a bit more to get explicit solutions for $g$.  The
equation of motion~(\ref{eomPLSA}) can be equivalently written as
$$
dgg^{-1}+(r_{+}-Ad_{g}r_{+}Ad_{g}^{-1})A=0
$$
and recalling that ${\tilde A}=Ad_{g}^{-1}A$,
$$
dgg^{-1}+r_{+}A-Ad_{g}r_{+}{\tilde A}=0
$$
$r_{+}$ projects on the $+$ component of the elements of ${\gl g}$ and
yields
$$
dgg^{-1}+h_{+}dh_{+}^{-1}-Ad_{g}{\tilde h}_{+}d{\tilde h}_{+}^{-1}=0
$$
which is equivalent to
\begin{equation} \label{ghat}
{\tilde h}_{+}^{-1}g^{-1}h_{+}d(h_{+}^{-1}g{\tilde h}_{+})=0
\Rightarrow h_{+}^{-1}g{\tilde h}_{+}={\hat g}\,,\qquad 
{\hat g}=g(\sigma_0) \in G
\end{equation}

The same procedure can be carried out by using $r_{-}$,
\begin{equation} \label{ghatprime}
h_{-}^{-1}g{\tilde h}_{-}={\hat g}
\end{equation}

The solutions for the equations of motion can then be expressed by the
single relation in $D$:
\begin{equation}
(g(\sigma){\tilde h}_+(\sigma),g(\sigma)\tilde h_-(\sigma))=
(h_+(\sigma)\hat g,h_-(\sigma)\hat g)
\end{equation}
which in particular implies that $(h_+(\sigma)\hat g,h_-(\sigma)\hat
g)\in D_0$, and $g(\sigma)$ takes values in the connected component of
orbits of $\hat g$ by dressing transformations.  These orbits are the
symplectic leaves of the Poisson-Lie group $G$.

Now we proceed to the hamiltonian analysis of the model. To that end
we fix the open geometry for $\Sigma=[0,\pi]\times \NR$ and take free
boundary conditions for $g$ and $A$ vanishing on vectors tangent to
$\partial\Sigma$. The boundary conditions enforce that $h_\pm, {\tilde
h}_\pm$ are constant along the connected components of the boundary,
$h_\pm(0,t)=h_{0\pm}$, ${\tilde h}_{\pm}(0,t)={\tilde h}_{0\pm}$,
$h_\pm(\pi,t)= h_{\pi\pm}$, ${\tilde h}_\pm(\pi,t)={\tilde
h}_{\pi\pm}$.  The presymplectic form is
$$
\omega=\int_\CC\tr(\delta g g^{-1}\wedge \delta g g^{-1}A-\delta g
g^{-1}\wedge \delta A)
$$
where $g$ and $A$ varies in the space of solutions of the equations of
motion and $\CC$ is any curve joining the two components of the
boundary. The result is, of course, independent of the particular
choice of $\CC$.

If we parametrize the solutions in terms of $h_\pm(\sigma), {\hat g}$
with $(h_+(\sigma)\hat g,h_-(\sigma)\hat g)\in D_0$ we obtain
\begin{equation}
\omega= \frac 12 \int_\CC d\Omega\left( (h_+(\sigma)\hat
g,h_-(\sigma)\hat g)\right)
\end{equation}
where $d$ acts on the \emph{space-time} variables in $\Sigma$.
Then $\omega$ depends only on the values at the boundaries, namely
$$
\omega= \frac 12 \left[\Omega\left((h_{\pi+}\hat g,h_{\pi-}\hat
g)\right)- \Omega((h_{0+}\hat g,h_{0-}\hat g))\right]
$$

Or if we take $\sigma_0=(0,t_0)$, i.e.\ $h_{0\pm}={\tilde h}_{0\pm}=e$
$$
\omega= \frac 12 \Omega\left((h_{\pi+}\hat g,h_{\pi-}\hat g)\right).
$$

From the analysis of the degeneracy of $\omega$ we conclude that the
vector fields induced by infinitesimal gauge transformations that are
the identity at the boundary form its kernel and we reduce the space
of solutions by this transformations.  The reduced phase space $P$ is
then the set of pairs $([(h_+,h_-)],\hat g)$ with $[(h_+,h_-)]$ a
homotopy class of maps from $[0,\pi]$ to $G_r$ which are the identity
at $0$ and have fixed value at $\pi$ and such that $(h_+(x)\hat
g,h_-(x)\hat g)\in D_0,\ x\in[0,\pi]$. The symplectic form on $P$ can
be viewed as the pullback of $\Omega$ by the map $([h_{+},h_{-}],{\hat
g})\mapsto (h_{\pi +}{\hat g},h_{\pi -}{\hat g})$.

The set of solutions may be endowed with the structure of a symplectic
groupoid (\cite{CaFe}). The product of two such solutions
$([(h_+,h_-)],\hat g)$ and $([(h'_+,h'_-)],\hat g')$ is defined if
$h_{\pi\pm}\hat g{\tilde h}_{\pi\pm}^{-1}=\hat g'$ and is given by
$([(h_+'',h_-'')],\hat g)$ with
\begin{equation} 
h''_\pm(x)=
\begin{cases}
h_\pm(2x) & 0\leq x\leq \frac12\pi
\\
h'(2x-\pi)_\pm h_{\pi\pm}&\frac 12\pi\leq x\leq \pi
\end{cases}
\end{equation} 

The results obtained in this section have their dual counterpart
if we consider the dual group $G^*$, which is accomplished 
in the next section. 

\section{Poisson-Lie Sigma model on $G^{*}$} 
\label{Gdual}\label{section5}

In order to unravel the role of the duality between Poisson-Lie groups
in the context of Poisson-Lie Sigma models we proceed now to the study
of the theory for $G^*\equiv G_r$.  Using the Poisson structure given
in~(\ref{dualPoissonLie}), the action of the model reads,
\begin{eqnarray}\label{dualPLS}
S_{\rm PL}^{*}(g_{+},g_{-},A)&=&\int_\Sigma
\tr\biggl[\left(d\gp\gpinv -d\gm\gminv\right)\wedge A+ 
\nonumber\\&&
\hphantom{\int_\Sigma\tr\biggl[}
+ \frac 12
A\wedge\left(\Ad_{\gp}-\Ad_{\gm}\right)\left(r_+\Ad_{\gm}^{-1}
-r_-\Ad_{\gp}^{-1}\right)A\biggr]
\end{eqnarray}

As shown in~\cite{ASS,Falceto}, this Poisson-Lie Sigma model with
target $G_{r}$ and fields $(g_{+},g_{-})$ and $A$ is closely related
to the $G/G$ gauged WZW model with fields $g=g_{-}g_{+}^{-1}$ and
$A$. We do not pursue further this relation in the present work. We
rather shall make a study analogous to that of the previous section,
stressing its similarities and differences with respect to the
Poisson-Lie Sigma model for $G$.

The equations of motion of the model are
\begin{subequations}
\begin{eqnarray}
g_{\pm}^{-1}dg_\pm+r_{\pm}(Ad_{g_+}^{-1}-Ad_{g_-}^{-1})A&=&0
\label{eomG*A}
\\
d{\tilde A}+[{\tilde A}, {\tilde A}]&=&0
\label{eomG*B}
\\
{\rm with}\ {\tilde A}:&=&(r_+ Ad_{g_-}^{-1}-r_- Ad_{g_+}^{-1})A\,.
\nonumber
\end{eqnarray}
\end{subequations}

\pagebreak[3]

Note that, similarly to the previous case for the group $G$, the
relation between $\tilde A$ and $A$ is nothing but the coadjoint
action of $(g_+,g_-)$ for the group $G_r$.  Like in the former case it
is posible to deduce, from the previous equations, a zero curvature
condition for $A$ itself
\begin{equation}\label{Afl}
dA+[A,A]=0\,.
\end{equation}

The gauge transformations, for $\beta:\Sigma\rightarrow{\gl g}$, read:
\begin{subequations}
\begin{eqnarray}
g_{\pm}^{-1}\delta_\beta g_\pm&=&r_{\pm}(Ad_{g_-}^{-1}-Ad_{g_+}^{-1})
\beta
\label{gaugeGsA}
\\
\delta_\beta A&=&d\beta+[A,\beta]+
\label{gaugeGsB}
\\  &&
{}+ 
\frac 12
\left(r_+Ad_{g_-}+r_-Ad_{g_+}\right)
\left[g_{+}^{-1}dg_+-g_{-}^{-1}dg_-+
\left(Ad_{g_+}^{-1}-Ad_{g_-}^{-1}\right)A,\tilde\beta\right]
\nonumber
\end{eqnarray}
\end{subequations}
where $\tilde\beta:=(r_+ Ad_{g_-}^{-1}-r_- Ad_{g_+}^{-1})\beta$.  The
change in the action is $\delta_\beta S_{\rm PL}^{*}=\int_\Sigma
d\,\tr((\gp d\gpinv -\gm d\gminv)\beta)$.  As before, if we forget
about the third term on the right hand side of~(\ref{gaugeGsB}), that
vanishes on-shell, the transformations close. Namely
$[\delta_\beta,\delta_\gamma]=\delta_{[\beta,\gamma]}$ which
corresponds to the gauge group $G$.

The solutions of the equations of motion can be obtained along
the same lines as before. From~(\ref{Afl})
\begin{equation}
A=hdh^{-1}
\end{equation}
in any contractible, open set. Without loss of generality we fix again
a point $\sigma_0\in\Sigma$ such that $h(\sigma_0)=e$. Now we write
together the equations for $g$ using the solution for $A$.
$$
g_{+}^{-1}dg_+
-g_{-}^{-1}dg_-+\left(Ad_{g_+}^{-1}-Ad_{g_-}^{-1}\right)hdh^{-1}=0
$$ 
or equivalently
$$ 
h^{-1}g_+g_-^{-1}h\, d(h^{-1}g_-g_+^{-1}h)=0\Rightarrow 
h^{-1}g_-g_+^{-1}h=\hat g_-\hat g_+^{-1}
$$
where $\hat g_\pm=g_\pm(\sigma_0)$. Hence
$$
g_+^{-1}h {\hat g}_+=g_-^{-1}h {\hat g}_-=:{\tilde h}
$$
and once more, we can obtain the solutions from an equation in $D$
$$
(g_+(\sigma){\tilde h}(\sigma)
,g_-(\sigma){\tilde h}(\sigma))=
\left({h}(\sigma)\hat g_+,
{h}(\sigma)\hat g_-\right),
$$
which means that $(g_+,g_-)$ is the dressing-transformed of $(\hat
g_+,\hat g_-)$ by $h$. Orbits of the dressing transformation are the
symplectic leaves of $G_r$.  Note, at this point, the complete
symmetry with the previous situation under the exchange of the roles
of $G$ and $G_r$.

We consider again the open geometry $\Sigma=[0,\pi]\times\NR$ with
boundary conditions for $A$ vanishing on vectors tangent to the
boundary and $g$ free. The boundary conditions impose that $h$ and
$\tilde h$ are constant in every connected component of
$\partial\Sigma$ i.e.\ $h(0,t)=h_0$, $h(\pi,t)=h_\pi$, $\tilde
h(0,t)=\tilde h_0$, $\tilde h(\pi,t)=\tilde h_\pi$.

The presymplectic form of the model $\omega^{*}$ can be written
\begin{equation}
\omega^{*}= \frac 12 \int_\CC d \Omega({h}(\sigma)\hat
g_+,{h}(\sigma)\hat g_-)
\end{equation}
where again $d$ acts on the $\sigma$ variables. Therefore
\begin{equation}
\omega^{*}= \frac 12 \left[\Omega\left(({h_\pi}\hat g_+,{h_\pi}\hat
  g_-)) -\Omega(({h_0}\hat g_+,{h_0}\hat g_-)\right)\right]
\end{equation}
and taking $\sigma_0=(0,t_0)$, or $h_0=\tilde h_0=e$,
\begin{equation}
\omega^{*}= \frac 12 \Omega\left(({h_\pi}\hat g_+,{h_\pi}\hat
g_-)\right).
\end{equation}

Points of the phase space reduced by gauge symmetries $P^*$, can be
described by pairs $([h],(\hat g_+,\hat g_-))$ where by $[h]$ we
denote a homotopy class of maps $h:[0,\pi]\rightarrow G$ with fixed
boundary values $h(0)=e$, $h(\pi)=h_\pi$ and such that $({h}(x)\hat
g_+, {h}(x)\hat g_-) \in D_0$. The symplectic form on $P^*$ can be
described as the pullback of $\Omega$ by the map $([h],(\hat g_+,\hat
g_-))\mapsto (h_\pi \hat g_+,h_\pi \hat g_-)$.

Note the duality between $P$ and $P^*$. The symplectic forms of the
two models coincide upon the exchange of $h_\pi$ with $\hat g^{-1}$
and $(\hat g_+,\hat g_-)$ with
$(h_{\pi+}^{-1},h_{\pi-}^{-1})$. Variables $(h_{\pi+},h_{\pi-})$ in
the $G$ model or $h_\pi$ in the $G^*$ model are associated to boundary
values of their corresponding fields (this can be more explicitly seen
working in a manifold $\Sigma$ whose boundary has more than two
connected components) while $\hat g$ and $(\hat g_+,\hat g_-)$ are
associated to the bulk. In this sense one can say that the duality
previously described between the Poisson-Lie Sigma models for $G$ and
$G^*$ can be stated as a bulk-boundary duality.

\section{More general Poisson Sigma models on $G$} 
\label{generalization}\label{section6}

In this section we solve the model with Poisson
structure~(\ref{generalPoisson}) following the lines of
section~\ref{section4}.  As remarked in section~\ref{section3} this
structure does not make $G$ a Poisson Lie group, but indeed the left
and right product by the corresponding Poisson-Lie groups are Poisson
actions.  In the resolution of the model we shall introduce a
generalization of the dressing transformation and of the Heissenberg
double and we shall be able to identify the symplectic leaves for the
Poissson structure defined on $G$.

The action for the model is,
\begin{equation} \label{PoissonSigma}
S(g,A)=\int_{\Sigma}\tr(dgg^{-1}\wedge A)- \frac 14 \tr(A\wedge
(r+Ad_{g}r'Ad_{g}^{-1})A)
\end{equation}
where as stressed in section~\ref{section3} $r$ and $r'$ are two
solutions of the modified Yang-Baxter equation~(\ref{YangBaxter}) with
the same value for $\alpha$.  The equations of motion are
\begin{eqnarray} 
dgg^{-1}+ \frac 12 \left(r+Ad_{g}r'Ad_{g}^{-1}\right)A&=&0 
\label{eomPoissonSigmaA}
\nonumber\\
d{\tilde A}-[{\tilde A},{\tilde A}]_{r'}&=&0\,,\qquad 
{\tilde A}:=Ad_{g}^{-1}A
\label{eomPoissonSigmaB}
\end{eqnarray}

From the previous equations, or performing in the action variations of
the fields that keep $\tilde A$ unchanged, we obtain
\begin{equation} 
\label{eomA2}
dA+[A,A]_{r}=0
\end{equation} 

The gauge symmetry in its infinitesimal form is
\begin{eqnarray} 
\delta_{\beta}gg^{-1}&=&- \frac 12 (r+Ad_g r'Ad_{g}^{-1})\beta 
\label{infsymmetrygeneralA}
\nonumber\\
\delta_{\beta}A&=&d\beta + [A,\beta]_{r}- \frac 12 \left[dgg^{-1}+
 \frac 12 \left(r+Ad_{g}r'Ad_{g}^{-1}\right)A,\beta\right]
\label{infsymmetrygeneralB}
\nonumber\\
&&\beta:\Sigma \rightarrow {\gl g}
\end{eqnarray}

The transformation for $g$ corresponds to the right \emph{dressing}
vector field that comes, like in the Poisson-Lie case of
section~\ref{section4}, from the contraction of the Poisson structure
with the right-invariant forms.  On-shell,
$[\delta_{\beta_{1}},\delta_{\beta_2}]=
\delta_{[\beta_{1},\beta_{2}]_{r}}$. Hence, the symmetry group is the
one corresponding to the matrix $r$, i.e.\ $G_{r}$.  The reason for
the preferred role of $r$ against $r'$ in the gauge symmetry is simply
the choice of right-invariant one forms to express the Poisson
structure.  Had we chosen left-invariant forms (i.e.\ changing the
variable $A$ by $\tilde A$ in the action) the symmetry algebra would
have been that of $r'$.

As before we shall take $\alpha=1$.  Then we can consider $G_r$ or
$G_{r'}$ as different subgroups of the same double group $D$.  We
shall denote by $(h_+,h_-)$ (resp. $(h_{+'},h_{-'})$) the elements of
$G_r$ (resp. $G_{r'}$).

Using the methods of section~\ref{section4}, we can easily solve the
model. As before, we first write locally the solutions for $A$ and
${\tilde A}$,
\begin{eqnarray} \label{solutionsAAtilde}
A&=&h_{+}dh_{+}^{-1}-h_{-}dh_{-}^{-1}
\nonumber\\
{\tilde A}&=&{\tilde h}_{-'}d{\tilde h}_{-'}^{-1}-
{\tilde h}_{+'}d{\tilde h}_{+'}^{-1}
\\
&&(h_{+},h_{-}):U \rightarrow G_{r}\,,\qquad 
({\tilde h}_{+'},\ {\tilde h}_{-'}):U \rightarrow G_{r'}\,,\qquad
\Sigma \supset U\ \textrm{open contract.}
\nonumber
\end{eqnarray}
with $h_\pm(\sigma_0)=\tilde h_{\pm'}(\sigma_0)=e$

The equation of motion for $g$ eq.~(\ref{eomPoissonSigmaA}) can be
transformed into
$$
dgg^{-1}+r_{\pm}A+Ad_{g}r'_{\mp}\tilde A=0\,,
$$
and  inserting the solutions for $A$, $\tilde A$
$$
dgg^{-1}+h_{\pm}dh_{\pm}^{-1}-
Ad_{g}{\tilde h}_{\mp'}d{\tilde h}_{\mp'}^{-1}=0
$$

Or equivalently
$$
h_+^{-1}g\tilde h_{-'}=h_-^{-1}g\tilde h_{+'}=\hat g
$$
with $\hat g=g(\sigma_0)\in G$.

We can write the general solution as an equation in $D$ for
$g(\sigma)$,
\begin{equation}
\label{solgen}
(g(\sigma)\tilde h_{-'}(\sigma),g(\sigma)\tilde
h_{+'}(\sigma))= (h_+(\sigma)\hat g,h_-(\sigma)\hat g)\,.
\end{equation}

If we now define $D'_0:= G_rG_d\cap G_dG_{-r'}$, we see that for
solutions of the equations of motion $(h_+(\sigma)\hat
g,h_-(\sigma)\hat g)\in D'_0$.  The symplectic leaves of $G_{r,r'}$
are connected components of the orbits of the generalized dressing
transformation of $\hat g$ by $(h_+,h_-)\in G_r$, that comes from
solving eq.~(\ref{solgen}) in $g(\sigma)$.

In order to describe the presymplectic structure in the space of
solutions we need to introduce a new Poisson bracket in $D$. Recall
that the Heisenberg double was defined in section~\ref{section3} as
$D_{R,R}$ with $R=P_d-P_r$. We can generalize this construction by
introducing another $r$-matrix $r'$ that give rise to
$R'=P_d-P_{-r'}$.  The Poisson structure in the double we are
interested in is $D_{R,R'}$ that is again non-degenerate around the
unit. Its main symplectic leaf is $D'_0$.  If we parametrize the
points in $D'_0$ by $(\eta_+\xi,\eta_-\xi)=
(\tilde\xi\tilde\eta_{-'},\tilde\xi\tilde\eta_{+'})$, the symplectic
structure in $D'_0$ obtained by inverting the Poisson bracket is
$$
\Omega'(\eta_+\xi,\eta_-\xi)= \tr\left[d\tilde\xi\tilde \xi^{-1}\wedge
\left(d\eta_{+}{\eta}_{+}^{-1}-d{\eta}_{-}{\eta}_{-}^{-1}\right)
-\xi^{-1}d \xi\wedge\left(\tilde \eta_{+'}^{-1}d\tilde \eta_{+'}-
\tilde \eta_{-'}^{-1}d\tilde \eta_{-'}\right)\right].
$$

Note again that although for a given point of $D'_0$ factors
$\xi,\eta_{\pm},\tilde\xi,\tilde\eta_{\pm'}$ in general are not
uniquely determined (different choices differ by elements of the
discrete groups $G_0$ or $G_0'$) the form $\Omega'$ is not affected by
the ambiguity and is indeed well defined in $D_0'$.
 
The presymplectic structure of the $r,r'$ Poisson-Sigma model
$\omega'$ in the open geometry ($\Sigma=[0,\pi]\times\NR$) with the
boundary conditions of section~\ref{section4} can then be written in
terms of $\Omega'$. It reads
$$
\omega'=\Omega'((h_{\pi+}\hat g,h_{\pi-}\hat g))- \Omega'((h_{0+}\hat
g,h_{0-}\hat g))
$$  
with $h_{0\pm}=h_\pm(0)$ $h_{\pi\pm}=h_\pm(\pi)$. And if we take
$\sigma_0=(0,t_0)$, i.e.\ $h_{0\pm}=e$,
$$
\omega'=\Omega'\left((h_{\pi+}\hat g,h_{\pi-}\hat g)\right).
$$

The discussion of the gauge transformations and the reduced phase
space goes parallel to the previous models. The points of the reduced
phase space $P'$ are pairs $([(h_+,h_-)],\hat g)$ with $[h_+,h_-]$ the
homotopy class of maps $[0,\pi]\rightarrow G_r$ with fixed boundary
values and such that $(h_{+}(x)\hat g,h_{-}(x)\hat g)\in D_0'$. The
symplectic form in $P'$ can be obtained as the pullback of $\Omega'$
by the map $([(h_+,h_-)],\hat g)\mapsto (h_{\pi+}\hat g,h_{\pi-}\hat
g)\in D_0'$.

\section{Conclusions} \label{conclusions}\label{section7}

We have solved the Poisson Sigma model for a Poisson-Lie group $G$ and
for its dual group $G^*$.  This models have non-regular
foliations. For example, as both are Poisson-Lie groups, the unit of
the respective groups forms itself a symplectic leaf.  Due to this
fact they fall out of the general construction of solutions
of~\cite{BojoStrobl} and, as we have the explicit solutions at hand,
they are an interesting example in which we can test the methods
of~\cite{BojoStrobl} for the case of non regular leaves.  Actually in
many aspects our models can be taken as a deformation of the linear
one.

Another remarkable aspect of the models under consideration is their
duality.  We can take $G$ and $G^*$ as subgroups of the same double
group $D=G\times G$, $G_d$ and $G_r$ respectively. In the resolution
of the two models on the strip $[0,\pi]\times\NR$, we find a close
relation between their reduced phase spaces (obtained by taking
quotient by the gauge symmetry).  Actually we see that there exist
maps from these spaces to $D_0$ such that their symplectic forms are
obtained as the pullback by these maps of the symplectic form on the
main symplectic leaf of the Heisenberg double. This study reveals a
bulk-boundary duality, such that the degrees of freedom naturally
associated to the bulk in one model are mapped into those
corresponding to the boundary in the other one.  This suggests a
connection with the Poisson-Lie non-abelian T-duality
of~\cite{Klimcik} for the case of WZW models. The concrete form of
this correspondence is not clear to us.

We have also solved the model on $G$ for the more general Poisson
structure associated to a pair of $r$ matrices introduced
in~\cite{STS}. The solution is obtained along similar lines to those
of the Poisson-Lie case. Now the symplectic form in the reduced phase
space is obtained from a new Poisson structure in the double that
generalizes the Heisenberg double of~\cite{STS,STS2,Alekseev}.

Throughout the paper we have considered $r$- matrices that are
solutions of the modified Yang-Baxter equation~(\ref{YangBaxter}) with
$\alpha=1$, i.e.\ factorizable Lie bialgebras of ref.~\cite{STS2}.  In
this case we can take the same double group $D=G\times G$
independently of the choice of $r$ because only $G_r$, the embedding
of $G^*$ in $D$, depends on it. This allows us to carry out the study
of the general case for a pair of $r$-matrices in
section~\ref{section6}.  The construction can be generalized without
difficulties if $\alpha\not=0$, but for the case $\alpha=0$ our
methods have to be modified and the explicit form of the solutions
requires further work.  Despite this fact we would expect that the
main results about duality remain essentially unchanged.

For instance, for the trivial model of eq.~(\ref{PLS}) with $r=0$ and
its dual, the linear one of eq.~(\ref{linac}), the reduced phase space
is $T^*(G)\cong G\times{\gl g}^{*} \cong G\times{\gl g}$.  In the
trivial model the bulk degrees of freedom are in $G$ and those of the
boundary in $\gl g$, whereas for the linear model the situation is the
converse.  The symplectic form in both cases is the canonical form in
$T^*(G)$.

As extensions of the work carried out in the paper one might consider
more general geometries and boundary conditions.  Another point that
is worth mentioning is that of possible non-topological deformations
of our models.  In order to keep the gauge symmetry of the model one
usually considers the addition of Cassimir functions on the Poisson
manifold. One of these deformations, for the linear model, gives rise
to the two dimensional Yang-Mills theory. It would be interesting to
consider a generalizations of the latter in the context of Poisson-Lie
groups.

An aspect that we do not cover in this paper is that of quantization
of the models. The cases for Poisson-Lie groups have been extensively
studied in the literature; see for
instance~\cite{STS2,Falceto,AF}. However, quantization of the
generalized Heisenberg double $D_{R,R'}$ of section~\ref{section6} is
much less known. It will be the subject of further research.

\end{document}